\newcommand{\sym}[3]{ 
    \num[round-mode=places, round-precision={#3}]{#1}
    \pm
    \num[round-mode=places, round-precision={#3}]{#2}
}
\newcommand{\asym}[4]{ 
    \num[round-mode=places, round-precision={#4}]{#1}
    ^{+\num[round-mode=places, round-precision={#4}]{#2}}
    _{-\num[round-mode=places, round-precision={#4}]{#3}}
}

\def\blgnot{BLG723.31.34651}
\def\blg{BLG723}

\def\msol{\rm M_\odot}
\def\rsol{\rm R_\odot}
\def\deg{\rm deg}
\def\kelvin{\rm K}
\def\kms{\rm km~s^{-1}}
\def\mas{\rm mas}

\def\vegamag{\rm Vega~mag}
\def\s{\rm s}
\def\min{\rm min}

\def\rhocgs{\rm g~cm^{-3}}
\def\myr{\rm Myr}

\def\sourceid{4099585591585929856}
\def\parallax{\sym{0.4415839359709078}{0.06358759}{3}}
\def\ra{18:43:24.18}
\def\dec{$-$17:01:41.25}

\def\gmag{\sym{16.439858736041153}{0.004861695463806766}{3}}

\def\bpmag{\sym{16.467746149033417}{0.012745895245571097}{2}}
\def\rpmag{\sym{16.28424661907662}{0.01614589385675299}{2}}
\def\agphot{0.7505}
\def\ebprp{0.4037}

\def\periodmin{63.8959988(21)}
\def\periods{3833.7599\pm0.0001}

\def\lcurveq{\asym{0.7795879312529591}{0.04219653985669347}{0.04261887288604471}{2}}
\def\lcurveiangle{\asym{79.82025859837141}{0.3962078065219288}{0.3534889372123331}{1}}
\def\lcurverwd{\asym{0.0362810778542771}{0.013766003186181701}{0.009935405839040304}{3}}
\def\lcurvetwd{\asym{12098.496105717746}{14406.293483962363}{8656.339301679265}{-1}}
\def\lcurvetsdb{\asym{35800.02138152307}{615.9553573632074}{620.6973459446235}{-1}}
\def\lcurvemidtime{2460936.622401~(\pm0.78)}
\def\lcurvevscale{\asym{592.6779023667533}{18.94249689040157}{18.000115782495413}{0}}

\def\lcurverdisc{\asym{0.3764724882685884}{0.0067546600314567495}{0.006404039847408338}{3}}
\def\lcurvehdisc{\asym{0.0514517106769978}{0.008784346252701199}{0.006948341348076099}{3}}
\def\lcurvetdisc{\asym{13857.953089900358}{776.0264218788943}{1023.9300141554322}{-1}}
\def\lcurvetempedge{\asym{4349.954055436972}{4150.185169517239}{2948.8926459944046}{-1}}

\def\lcurveslopeB{\asym{0.0011131490309064}{0.0010815650665120002}{0.0010926781159297}{4}}
\def\lcurveldcAB{\asym{0.0549474514524565}{0.0016822984142881997}{0.001735610862466895}{3}}
\def\lcurveldcBB{\asym{0.2604158560776771}{0.006115029205869704}{0.006555702258890683}{3}}
\def\lcurvegdcB{\asym{0.384693783231037}{0.014147306832603512}{0.01369369387167868}{3}}

\def\lcurveslopeR{\asym{0.0072874135003333}{0.0013322225113222997}{0.0012844907932558006}{4}}
\def\lcurveldcAR{\asym{0.04008134251115245}{0.001417411285104947}{0.0013882813695360527}{3}}
\def\lcurveldcBR{\asym{0.180212650979722}{0.005307727793164491}{0.0053790086026453965}{3}}
\def\lcurvegdcR{\asym{0.3799885753777695}{0.015524688304340728}{0.015095826905733856}{3}}

\def\lcurveslopeg{\asym{-0.0043220551684787}{0.001610899181967}{0.0016533371736071533}{4}}
\def\lcurveldcAg{\asym{0.0515499868478883}{0.0017501236829926964}{0.0017978096670794863}{3}}
\def\lcurveldcBg{\asym{0.2445605314688228}{0.005764454799877111}{0.006029953399514398}{3}}
\def\lcurvegdcg{\asym{0.3796728909365912}{0.0162911567803683}{0.014224281241921655}{3}}

\def\lcurveslopei{\asym{-0.0097781161131658}{0.0007521432086096003}{0.000731550426365201}{4}}
\def\lcurveldcAi{\asym{0.040801529115458}{0.0011450735760588013}{0.0011152410358660006}{3}}
\def\lcurveldcBi{\asym{0.1708673309969054}{0.005179858756437089}{0.00479294799388269}{3}}
\def\lcurvegdci{\asym{0.40066518162308384}{0.011953654117236368}{0.011644127742292565}{3}}

\def\lcurvemb{\asym{0.4195807883983086}{0.05181736720292851}{0.04671514627187234}{2}}
\def\lcurvema{\asym{0.5384509211794289}{0.04447217030181505}{0.04012539161650791}{2}}
\def\lcurvea{\asym{0.5198065851854727}{0.01661346674840769}{0.015786953892625744}{2}}
\def\lcurvervol{\asym{0.18611187261430467}{0.007753557060490773}{0.0076039359196941325}{3}}

\def\lcurveqB{\asym{0.8409715244356485}{0.03759480077572752}{0.03748576324743924}{2}}
\def\lcurveiangleB{\asym{79.6090093295794}{0.3849793742953693}{0.3791071189281894}{1}}
\def\lcurverwdB{\asym{0.0324012906929894}{0.013776562376203096}{0.008491640156189934}{3}}
\def\lcurvetwdB{\asym{15905.469317840232}{14193.401087093329}{10824.331650069704}{-1}}
\def\lcurvetsdbB{\asym{35851.87827431634}{674.3001196694386}{611.1292159375589}{-1}}
\def\lcurvemidtimeB{2460936.622399~(\pm0.78)}
\def\lcurvevscaleB{\asym{626.1896604457737}{16.036110081765855}{13.649086934713637}{0}}

\def\lcurverdiscB{\asym{0.3826452685630895}{0.006643618842534227}{0.007643440419257075}{3}}
\def\lcurvehdiscB{\asym{0.0524252924070921}{0.0102764814997637}{0.006382961268326999}{3}}
\def\lcurvetdiscB{\asym{14245.29435449611}{726.4809042358083}{1113.8784190625738}{-1}}
\def\lcurvetempedgeB{\asym{5894.31566166531}{4917.361028029341}{4030.279925829322}{-1}}

\def\lcurveslopeBB{\asym{0.0008784993938642}{0.0010451786085992}{0.001094138243223998}{4}}
\def\lcurveldcABB{\asym{0.054943336212405}{0.0017179638404682188}{0.0017564140512712975}{3}}
\def\lcurveldcBBB{\asym{0.2608465607661643}{0.005456577185173561}{0.005711434819241079}{3}}
\def\lcurvegdcBB{\asym{0.3873585637124314}{0.016072404477055102}{0.015084960810020476}{3}}

\def\lcurveslopeRB{\asym{0.0074520521811436}{0.0013666448621730007}{0.0014011984731515}{4}}
\def\lcurveldcARB{\asym{0.0401235863268519}{0.0013745483547966952}{0.0014813703278461002}{3}}
\def\lcurveldcBRB{\asym{0.1802200675193269}{0.005334011250507309}{0.005090137495821628}{3}}
\def\lcurvegdcRB{\asym{0.3835809706577698}{0.014315367632571907}{0.015393813574176796}{3}}

\def\lcurveslopegB{\asym{-0.0050273413731879}{0.0014682610336185001}{0.0015068316902725797}{4}}
\def\lcurveldcAgB{\asym{0.0514496333627885}{0.0017690010035718781}{0.0016341655155520801}{3}}
\def\lcurveldcBgB{\asym{0.2446764159117534}{0.006259299760396714}{0.005831172457511996}{3}}
\def\lcurvegdcgB{\asym{0.3809319216375667}{0.01366926480332864}{0.01377179655771349}{3}}

\def\lcurveslopeiB{\asym{-0.0097306388306953}{0.0007703973975806012}{0.0007087991071833988}{4}}
\def\lcurveldcAiB{\asym{0.0407912316368124}{0.0012244967303607981}{0.0011290628860171}{3}}
\def\lcurveldcBiB{\asym{0.1710547380136811}{0.005711483379787496}{0.0060164971160937}{3}}
\def\lcurvegdciB{\asym{0.4070277733520518}{0.012781231930287962}{0.012486967582774233}{3}}

\def\lcurvembB{\asym{0.5163915654040682}{0.04922053279014427}{0.04112581362354745}{2}}
\def\lcurvemaB{\asym{0.6150388552808029}{0.03882002177162147}{0.03527332539925809}{2}}
\def\lcurveaB{\asym{0.5491979838879637}{0.01406442789240836}{0.011970895560811323}{2}}
\def\lcurvervolB{\asym{0.20046400334368755}{0.00660522061062091}{0.006124339006745305}{3}}

\def\sedmb{\asym{0.44}{0.22}{0.15}{2}}
\def\sedrvol{\asym{0.208}{0.034}{0.026}{3}}

\def\speclogg{\sym{5.531487874019269}{0.09257282561066647}{2}}
\def\specteff{\sym{35840.55489819961}{639.4389395660154}{-1}}
\def\spechelium{\sym{-1.0094108962367174}{0.021272929227981952}{2}}
\def\specvsini{\asym{244}{24}{10}{0}}
\def\speckb{\sym{328.68}{7.19}{0}}

\def\eggletondensity{\sym{91.99872}{2.01635}{1}}
\def\eggletondensityB{\sym{90.94866}{1.95576}{1}}
\def\eggletonmass{\sym{0.5876336}{0.0969217}{2}}

\def\pdot{\sym{-1.19274}{1.14416}{2}}
\def\pdotB{\sym{-8.96826}{1.12116}{2}}

\def\pdotoc{\sym{-1.44}{1.53}{2}}
\def\pdotoc{\sym{-3.11}{2.11}{2}}

\documentclass[apj, twocolumn]{openjournal}

\usepackage{siunitx}
\usepackage{amsmath}
\usepackage{graphicx}

\usepackage{float}
\usepackage{xcolor}
\usepackage{textgreek}
\usepackage[utf8]{inputenc}
\usepackage[english]{babel}

\usepackage{color,colortbl}
\definecolor{linkcolor}{rgb}{0.0,0.3,0.5}

\usepackage{hyperref}
\hypersetup{
    unicode, 
    colorlinks=true,
    linkcolor=linkcolor,
    citecolor=linkcolor,
    filecolor=linkcolor,
    urlcolor=linkcolor,
}
\usepackage{orcidlink}

\begin{document}
\title{\blgnot\ as an eclipsing Roche lobe-filling sdOB+WD binary}
\author{Alekzander Kosakowski$^1$\orcidlink{0000-0002-9878-1647}, Charles E. Derbyshire$^1$\orcidlink{0009-0006-1551-4225}, Sydney Jetton$^1$\orcidlink{0009-0001-6919-1533}, and Brad N. Barlow$^1$\orcidlink{0000-0002-8558-4353}}
\affiliation{$^1$Department of Physics and Astronomy, University of North Carolina at Chapel Hill, Chapel Hill, NC 27599, USA}

\begin{abstract}
    We present a detailed observational analysis of \blgnot, the fourth reported eclipsing member of the class of Roche lobe-filling hot subdwarfs, using time series spectroscopy and multi-band photometric observations from the Goodman High Throughput Spectrograph on the 4.1-m SOAR telescope. Our light curves show periodic photometric variability at $P=64~{\rm min}$ from the tidally-deformed hot subdwarf donor and deep eclipses of the donor by an opaque accretion disc. We report two light curve model solutions that agree with each other to within $2\sigma$, both suggesting that \blgnot\ contains a typical hot subdwarf donor, a white dwarf accretor, and an irradiated opaque accretion disc. Our spectroscopic analysis identifies statistically significant phase-dependent variations in the effective temperature and surface gravity coinciding with the phase-dependent variations seen in our observed light curves. We attribute these variations to a combination of intrinsic variations in local atmospheric parameters due to tidal deformation of the donor and unmodeled spectroscopic flux contributions from an accretion disc.
    \end{abstract}

\begin{keywords}
    {hot subdwarfs, close binaries, accreting binaries, white dwarfs}
\end{keywords}


\section{Introduction}
\label{sec:intro}

Hot subdwarfs (spectral class sdB, sdOB, sdO) are a class of helium-burning hot sub-luminous stars with stripped hydrogen envelopes and typical masses around $M_{\rm sd}\approx0.47~\msol$ \citep{heber1986, heber2009, heber2016, heber2025}. Hot subdwarfs are found along the extreme horizontal branch of the Hertzsprung-Russell diagram and experience relatively short helium-burning lifetimes ($\tau\approx100~{\rm Myr}$) before evolving to become white dwarfs.

Observationally, a large fraction of hot subdwarfs are found to be in detached close binaries with white dwarf, main sequence, and brown dwarf companions and with orbital periods $P_{\rm orb}=\mathcal{O}(10^{-2}-10^{1}~{\rm d})$ \citep{maxted2001, napiwotzki2004, copperwheat2011, kupfer2015, schaffenroth2022, he2025}. Their relatively high close binary fraction and thin hydrogen envelope suggest that their formation is dominated by binary interactions, either through early binary evolution stripping a giant star's hydrogen envelope or through late binary evolution as a possible outcome to the merger of two low mass white dwarfs \citep{webbink1984, han2002, han2003, hall2016}. In support of this, \citet{pelisoli2020} have demonstrated observationally that binary interaction may even be necessary for the formation of hot subdwarf stars.

Hot subdwarf stars may show periodic photometric variability caused by combinations of pulsations, tidal deformation, relativistic beaming, eclipses, and reprocessed irradiation from their close companions. Their blue colors and periodic photometric variability make hot subdwarfs relatively easy to identify in archival data from time domain surveys such as the Zwicky Transient Facility \citep[ZTF;][]{bellm2019a,graham2019,masci2019}, the Gravitational Lensing Experiment \citep[OGLE;][]{udalski2015}, and the Asteroid Terrestrial-impact Last Alert System \citep[ATLAS;][]{tonry2018,heinze2018}.

\begin{table*}[]
    \tablewidth{\textwidth}
    \small
    \centering
    \caption{\small Summary of our follow-up observations taken of \blg\ using the 4.1-meter SOAR telescope.}
    \begin{tabular*}{0.99\textwidth}{@{\extracolsep{\fill}}c c c c c}
    \hline
       Date & Instrument & Observation Type & Grating/Filter & Sequence Time\\
        {(UT)} & {} & {} & {} & {(minutes)}\\
        \hline
        \hline
       2025-06-20  &  Goodman Red HTS & Spectroscopy & $930~{\rm l~mm^{-1}}$ & 64\\
       2025-06-20  &  Goodman Red HTS & Photometry & Bessel $B$ & 74\\
       2025-06-20  &  Goodman Red HTS & Photometry & Bessel $R$ & 75\\
       2025-09-18  &  Goodman Red HTS & Photometry & SDSS $g'$ & 45\\
       2025-10-17  &  Goodman Red HTS & Photometry & SDSS $i'$ & 123\\
        \hline
    \end{tabular*}
    \label{tab:observations}
\end{table*}

The long-term monitoring of photometrically-variable compact binaries allows for indirect measurement of orbital decay caused by the emission of gravitational waves, which provides constraints to the mass of the binary \citep[see][]{hermes2012, teckenburg2025}. However, slow orbital decay rates ($\dot{P}\lesssim\mathcal{O}(10^{-13}-10^{-11}$)) and relatively short helium-burning lifetimes make it difficult for hot subdwarfs in binaries to begin mass transfer before evolving into white dwarfs. It is expected that hot subdwarfs which emerge from a common envelope with a white dwarf companion with orbital period $P_{\rm orb}\lesssim2-3~{\rm h}$ will begin mass transfer while burning helium \citep{bauer2021}. These mass-transferring systems may be identified photometrically as Roche lobe-filling hot subdwarfs \citep{kupfer2020b}.

The first two detected Roche lobe-filling hot subdwarfs, ZTF J2130+4420 \citep{kupfer2020a} and ZTF J2055+4651 \citep{kupfer2020b}, were discovered as a part of ZTF's high-cadence Galactic plane survey \citep{bellm2019b}, with orbital periods of only $P_{\rm orb}=39.3~{\rm min}$ and $P_{\rm orb}=56.3~{\rm min}$, respectively. These systems show periodic light curve variations from a tidally-deformed sdOB donor, deep eclipses of the bright sdOB donor by an opaque accretion disc, shallow eclipses of the compact white dwarf accretor by the hot subdwarf donor, and reprocessed irradiation from the outer accretion disc. Notably, neither of these systems shows direct evidence for flux contribution from the accretion disc itself \citep{deshmukh2023}.

The third member of this class of Roche lobe-filling hot subdwarfs, SMSS J1920-2001 \citep{li2022}, was discovered in archival data from the SkyMapper Southern Survey \citep[SMSS;][]{wolf2018, onken2019}. SMSS J1920-2001 was found to be an sdO-type hot subdwarf in an eclipsing accreting binary with a white dwarf at orbital period $P_{\rm orb}=209.90~{\rm min}$, but with poorly-constrained system parameters from their light curve modeling due to their lack of light curve models with accretion disc geometry included.

The fourth reported Roche lobe-filling hot subdwarf, ZTF J0007$+$4804 \citep{stringer2026}, has an orbital period $P_{\rm orb}=108.7~{\rm min}$ but does not show eclipses by an opaque accretion disc. The presence of its accretion disc was inferred through its regular dwarf nova outbursts seen in the archival light curves from sky surveys such as ZTF and TESS, with outburst recurrence time estimated to be $P_{\rm outburst}=9.0^{+0.7}_{-0.5}~{\rm d}$.

Here we present a detailed analysis of time series follow-up observations of \blgnot\ (hereafter \blg), the fifth reported member, and fourth eclipsing member, to the class of Roche lobe-filling hot subdwarfs. \blg\ was originally identified as a short-period variable star with orbital period $P_{\rm orb}=\periodmin~\min$ in archival data from the OGLE-IV project by \citet{borowicz2023}. The authors highlighted \blg\ as a likely Roche lobe-filling hot subdwarf in a binary with a white dwarf based on its distinct light curve shape and short photometric period. 

This manuscript is organized as follows: We describe our observations and data reduction in Sections \ref{sec:observations} and \ref{sec:data_reduction}. We provide the details of our spectroscopic and photometric analyses in Sections \ref{sec:spectro_analysis} and \ref{sec:photo_analysis}. Supplemental figures are presented in the appendix. For this work, we adopt the convention that the massive component (the accretor) is the primary star such that $M_1>M_2$ and that the bright hot subdwarf donor is eclipsed at orbital phase $\phi=0.5$ (secondary eclipse).

\section{Observations}\label{sec:observations}
\subsection{Spectroscopic observations}
We obtained 21 consecutive exposures of \blg\ using the Southern Astrophysical Research (SOAR) 4.1-meter telescope with the Goodman High Throughput Spectrograph \citep{clemens2004} Red Camera on UT 2025 June 20. Our configuration made use of the $930~{\rm lines~mm^{-1}}$ grating and $1.0~{\rm arcsec}$ long slit, providing spectroscopic resolving power $R\approx2000$ over the wavelength range $3600-5300~{\rm \AA}$. We aligned the slit position angle to include both \blg\ and a nearby field star to correct for instrument flexure and serve as a radial velocity reference.

Our observations covered one full binary orbit with 3-minute exposure time, resulting in minimal orbital smearing over the 64-minute binary orbital period. We obtained one set of three FeAr arclamp exposures at the end of our observing sequence, which were used for wavelength calibration.


\newpage
\subsection{Photometric observations}
We obtained four sets of time series photometry data of \blg\ using the Goodman High Throughput Spectrograph Red Camera on the SOAR 4.1-meter telescope. Our observations made use of Imaging mode with 2x2 binning. These observations are summarized in Table \ref{tab:observations}.

We obtained 73.7 minutes of time series photometry using the Bessel $B$ filter with 12-second exposures on UT 2025 June 20. Our observations spanned air mass between $1.16-1.44$. A fraction of this data was lost due to tertiary mirror instabilities, resulting in incomplete orbital phase coverage.

We obtained 74.9 minutes of time series photometry using the Bessel $R$ filter with 12-second exposures on UT 2025 June 20, immediately after our Bessel $B$ observations. Our observations spanned air mass between $1.44-2.13$. A small fraction of this data was lost due to mirror instabilities, resulting in incomplete orbital phase coverage.

We obtained 44.9 minutes of time series photometry using the SDSS $g'$ filter with 14-second exposures on UT 2025 September 18. The weather was clear with seeing between 1.0 and 1.5 arcsec, resulting in excellent image quality. These observations do not cover one full binary orbit, but do cover both eclipse features at signal-to-noise ratio $S/N>100$. These observations spanned air mass $1.27-1.48$.

We obtained 123.3 minutes of time series photometry using the SDSS $i'$ filter with 15-second exposures on UT 2025 October 18. The weather was clear with seeing at approximately $1.0~{\rm arcsec}$. These observations spanned air mass $1.21-2.11$.

\begin{figure}
    \centering
    \includegraphics[width=1.0\linewidth]{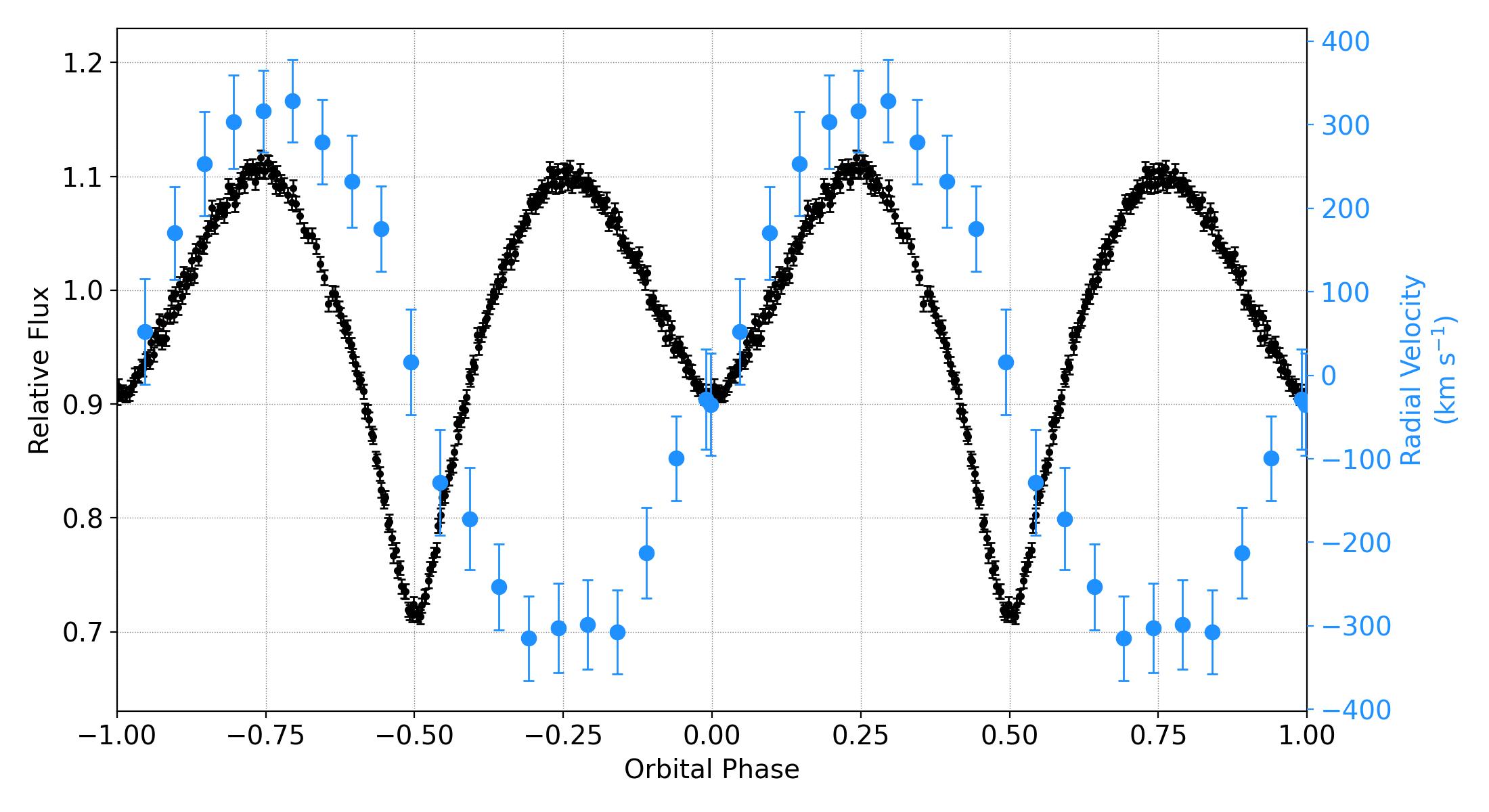}
    \caption{\small Phase-folded SDSS $i'$ light curve data of \blg\ (black). We over plot our measured radial velocity values from our optical spectroscopy in blue and using the right-hand axis. The radial velocity curve was artificially shifted to zero systemic velocity as part of our analysis. Data has been repeated over one orbit for visual clarity.}
    \label{fig:phase_combined}
\end{figure}

 \begin{table}
    \renewcommand{\arraystretch}{1.5}
    \small
    \centering
    \caption{\small Archival, Measured, and Derived parameters describing \blg.}
    \label{tab:main_table}
    
    \begin{tabular*}{0.99\columnwidth}{@{\extracolsep{\fill}}l r l}
        \hline\hline\\[-0.4cm]
        
        \multicolumn{3}{c}{Archival Parameters} \\
        \hline
            source\_id & $\sourceid$ & Gaia DR3\\
            ${\rm R.A.}$ & \ra & 2016\\
            ${\rm Dec.}$ & \dec & 2016\\
            Parallax & $\parallax$ & $\mas$\\
            ${\rm Gaia~G}$ & $\gmag$ & {\vegamag}\\
            ${\rm Gaia~G_{BP}}$ & $\bpmag$ & {\vegamag}\\
            ${\rm Gaia~G_{RP}}$ & $\rpmag$ & {\vegamag}\\
            $A_{\rm G}$ & $\agphot$ & {\vegamag}\\
            $E{\rm (G_{BP}-G_{RP})}$ & $\ebprp$ & {\vegamag}\\
            $P_{\rm orb,~OGLE}$ & $\periods$ & {$\s$}\\[0.3cm]
        
        \multicolumn{3}{c}{Optical Spectroscopy}\\
        \hline
            $\log\left[{\frac{\rm He}{\rm H}}\right]$ & $\spechelium$ & {} \\
            $\log{g_{\rm sdOB}}$ & $\speclogg$ & (cgs)\\
            $T_{\rm eff,~sdOB}$ & $\specteff$ & $\kelvin$ \\
            $v_{\rm sdOB}\sin{i}$ & $\specvsini$ & $\kms$ \\
            $K_{\rm sdOB}$ & $\speckb$ & $\kms$ \\[0.3cm]
                
        \multicolumn{3}{c}{Spectral Energy Distribution} \\
        \hline
            $R_{\rm sdOB}$ & $\sedrvol$ & $\rsol$ \\
            $M_{\rm sdOB}$ & $\sedmb$ & $\msol$ \\[0.1cm]

        \hline
        \hline
    \end{tabular*}
\end{table}

\newpage

\section{Data reduction}\label{sec:data_reduction}
\subsection{Spectroscopic data reduction}

We reduced our spectroscopic data using the Multi-Instrument Data Input Reducer\footnote{\url{https://github.com/Fabmat1/MIDIR}} (MIDIR) software package, which automates the data reduction process, including bias and flat field corrections. Spectroscopic wavelength solutions are determined by MIDIR through a Markov Chain Monte Carlo fitting approach to comparison lamp spectra and applied to each science spectrum individually.

\subsection{Photometric data reduction}
We reduced our time series photometry using standard CCD photometry procedures with custom Python software, including applying bias and flat corrections using calibration data obtained on the same nights as our observations.

We extracted the photometry of \blg\ and a set of similarly bright and non-variable field stars through an aperture photometry algorithm using the \texttt{photutils} package within \texttt{astropy}. We applied a centroid algorithm to center the apertures on the selected stars and used a constant aperture size for each dataset, chosen to minimize the flux error bar size in the final calibrated light curve.

We created a weighted-mean light curve from our selected non-variable field stars and divided this into the light curve of \blg\ to remove atmospheric effects. We did not attempt to apply a correction for color-dependent air mass effects, which are instead handled through our light curve modeling described in Section \ref{sec:photo_analysis}.

Finally, we inflated the error bar size of each of our photometric data sets by adding 0.5\% flux uncertainty in quadrature to account for unmeasured systematics, such as disc flickering effects, which would appear as relatively low-amplitude stochastic variations if present. We present the calibrated SDSS $i'$ light curve of \blg\ in Figure \ref{fig:phase_combined}.

\section{Spectroscopic analysis}\label{sec:spectro_analysis}

\subsection{Radial velocity measurements}

We measured the radial velocity of \blg\ from each spectrum by performing Gaussian fits to the individual hydrogen absorption lines ${\rm H\beta}$, ${\rm H\gamma}$, ${\rm H\delta}$, and ${\rm H\epsilon}$. Radial velocities were corrected for flexure using a cross-correlation technique against the same absorption lines from a comparison star placed within the same slit exposure.

We fit the radial velocity data of \blg\ using a sine curve with orbital period fixed at the OGLE period. We first fit for phase offset and systemic velocity. We then subtracted the systemic velocity and performed another sine curve fit for the velocity semi-amplitude. We find best-fitting velocity semi-amplitude $K=\speckb~\kms$. We present our individual phase-resolved radial velocity measurements of \blg\ in Figure \ref{fig:phase_combined} as blue points and using the right-hand axis.

\subsection{Phase-resolved atmospheric parameters}

We applied a $\chi^2$ minimization algorithm following the methods described in \citet{irrgang2014} to estimate the spectroscopic atmospheric parameters of \blg: effective temperature ($T_{\rm eff}$), surface gravity ($\log{g}$), helium abundance ($\log{\left[\frac{He}{H}\right]}$), and projected rotational velocity ($v\sin{i}$).

\begin{figure}
    \centering
    \includegraphics[width=1.0\linewidth]{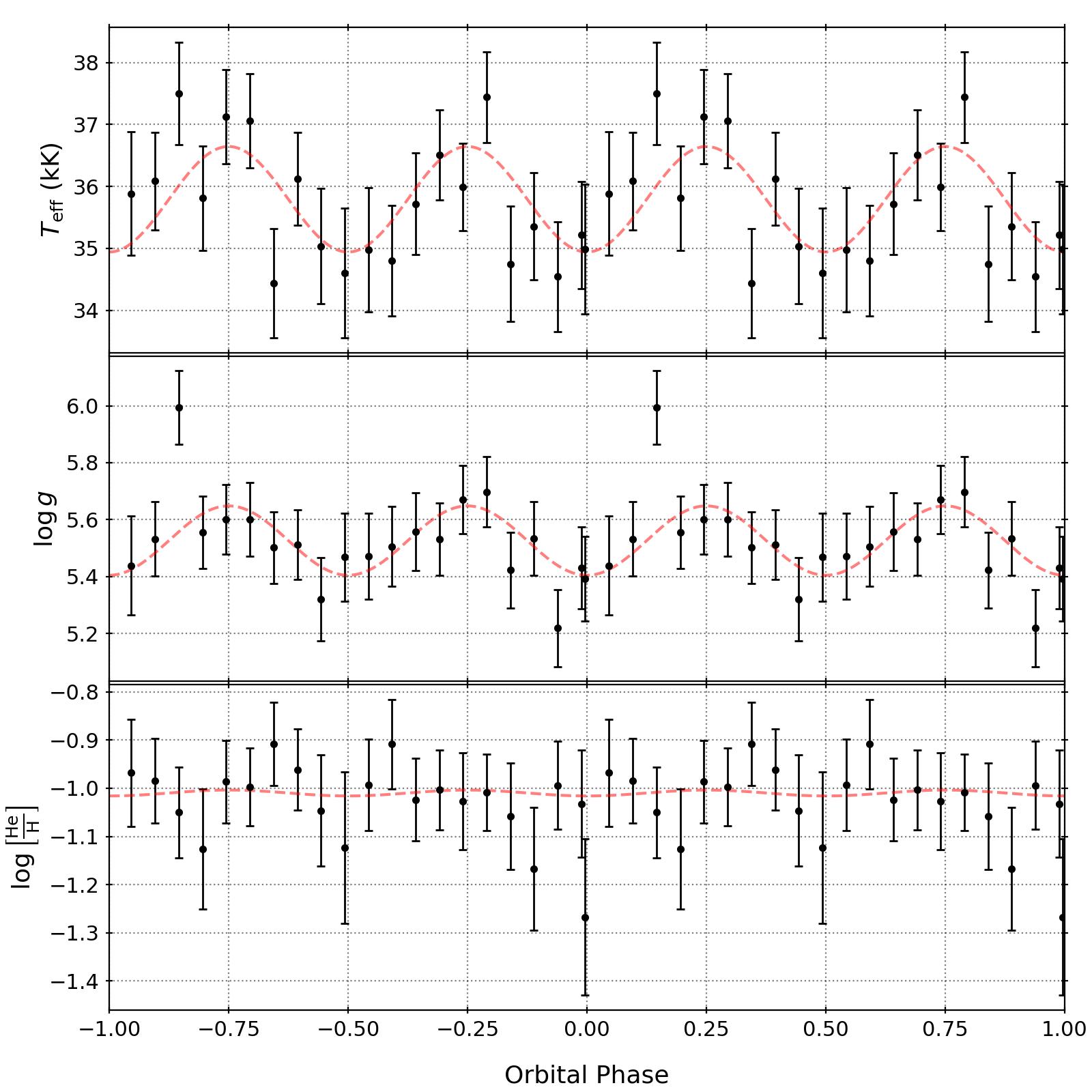}
    \caption{\small Best fitting atmospheric parameters to the individual spectra of \blg\ phase folded at the binary orbital period. Red dashed lines show the best-fitting sine curve to each fitted atmospheric parameter, using a fixed frequency and phase offset. Data has been repeated over one orbit for visual clarity.}
    \label{fig:timespec}
\end{figure}

We fit all 21 spectra simultaneously for individual effective temperature, surface gravity, and helium abundance, but required that all 21 spectra share the same fitted projected rotational velocity value at each fit iteration. Figure \ref{fig:timespec} shows the individual results from our spectroscopic atmospheric modeling plotted against orbital phase. We provide a representative best-fitting model spectrum over plotted onto our single-exposure observed spectrum of \blg\ at orbital phase $\phi=0.0$ in Appendix A Figure \ref{fig:spectrum_fit00}. We additionally include the spectroscopic fit to the summed spectrum of \blg\ in Figure \ref{fig:spectrum_fit} for reference. Notably, the absorption feature of He II $4686~{\rm \AA}$ is strongest near quadrature and lost to the noise of our single exposure spectra at other orbital phases.

Our phase-resolved spectroscopic modeling results show phase-dependent variations in both effective temperature and surface gravity. These variations appear to align in orbital phase with the key features seen in our light curves, suggesting that there may be a measurable difference in the spectroscopic atmospheric parameters caused by the changing visible surface of the tidally-deformed hot subdwarf donor star throughout its orbit. We observed maximum temperature and surface gravity at quadrature near orbital phase $\phi=0.25,~0.75$, aligning with the visibility of the He II $4686~{\rm \AA}$ absorption feature and with the light curve flux maxima. Similarly, minimum temperature and surface gravity align with light curve flux minima near orbital phase $\phi=0.00,0.50$, corresponding to mid-eclipse orbital alignment. 

We estimated the amplitude of variability seen in the atmospheric parameters of \blg\ by performing a least-squares fit to our phase-resolved spectroscopic fit results using a cosine function with fixed frequency and phase offset, retaining two free parameters for mean value and amplitude. We find the following best-fitting model values: $\left<\log{g}\right>=5.53\pm0.03$, $\left<T_{\rm eff}\right>=35790\pm190~{\rm K}$, and $\left<\log{\left[\frac{He}{H}\right]}\right>=-1.01\pm0.02$, with amplitudes $A_{\rm \log{g}}=0.12\pm0.04$, $A_{T_{\rm eff}}=850\pm260~{\rm K}$, and $A_{\log{\left[\frac{He}{H}\right]}}=0.01\pm0.03$.

We estimated the significance of these variations by performing an F-test comparing our cosine fit (two free parameters) against the weighted mean value (one free parameter) and find that the cosine model is the preferred model to 99.9\% confidence for surface gravity, 99.7\% confidence for effective temperature, and 15.8\% confidence for helium abundance. Similarly, for the effective temperature and surface gravity, we further compared the cosine model and weighted mean value model using the corrected Akaike Information Criterion \citep[AIC;][]{akaike1974, hurvich1989} and Bayesian Information Criterion \citep[BIC;][]{schwarz1978}. We find $\Delta{\rm AIC_{\log{g}}}=6.2$ and $\Delta{\rm BIC_{\log{g}}}=5.7$ and $\Delta{\rm AIC_{T_{eff}}}=8.4$ and $\Delta{\rm BIC_{T_{eff}}}=7.8$, statistically favoring the cosine model in both cases.

Variations in measured surface gravity and effective temperature with orbital phase are expected for binaries containing tidally-deformed stars. The observed amplitude of these variations depends on the magnitude of deformation and the orbital inclination of the binary. While these variations are typically lost to the noise in spectroscopic model fits, or ignored for simplicity, a sufficiently high signal-to-noise spectroscopic dataset with a dense atmospheric model grid may allow for the detection of these variations and provide an estimate to the mass ratio of the binary through Roche geometry constraints, degenerate with orbital inclination, further augmenting a detailed light curve analysis.

We used \texttt{lcurve} \citep{copperwheat2010} to estimate the expected amplitude of variability in surface gravity caused by tidal deformation in \blg\ by calculating the flux-weighted mean surface gravity over the visible surface of the donor throughout its orbit. We used system parameters derived from our light curve modeling, discussed in Section \ref{sec:lcurve}, to estimate the flux and surface gravity of the donor for these calculations. We find that the expected amplitude of surface gravity variations caused by tidal deformation is $A_{\rm \log{g}}=0.02~{\rm dex}$ when considering contributions from full visible stellar surface; significantly less than our measured amplitude for \blg. We expect that this discrepancy is caused by non-insignificant flux contributions from the accretion disc distorting the spectroscopic line profiles over the binary's orbit and confusing single-component spectroscopic modeling.

Given that our results for the surface gravity of \blg\ show significant phase-dependent variations on the level of $0.1~{\rm dex}$, providing an estimate of the true mass of the donor becomes complicated, since the donor mass is typically derived from the spectroscopic surface gravity which is assumed to have a singular value. For simplicity in this work, we treated the measured atmospheric parameter variations as a source of systematic error, rather than intrinsic variability, and used their weighted-mean values added in quadrature to the root-mean-squares of the cosine fit amplitudes ($\frac{A}{\sqrt{2}}$) to obtain our representative atmospheric parameter values for \blg, $\log{g}=\speclogg$, $T_{\rm eff}=\specteff~\kelvin$, and $\log{\left[\frac{He}{H}\right]}=\spechelium$, which we summarized in Table \ref{tab:main_table}.

\section{Photometric analysis}\label{sec:photo_analysis}

\subsection{Spectral energy distribution}
\begin{figure}
    \centering
    \includegraphics[width=1.0\linewidth]{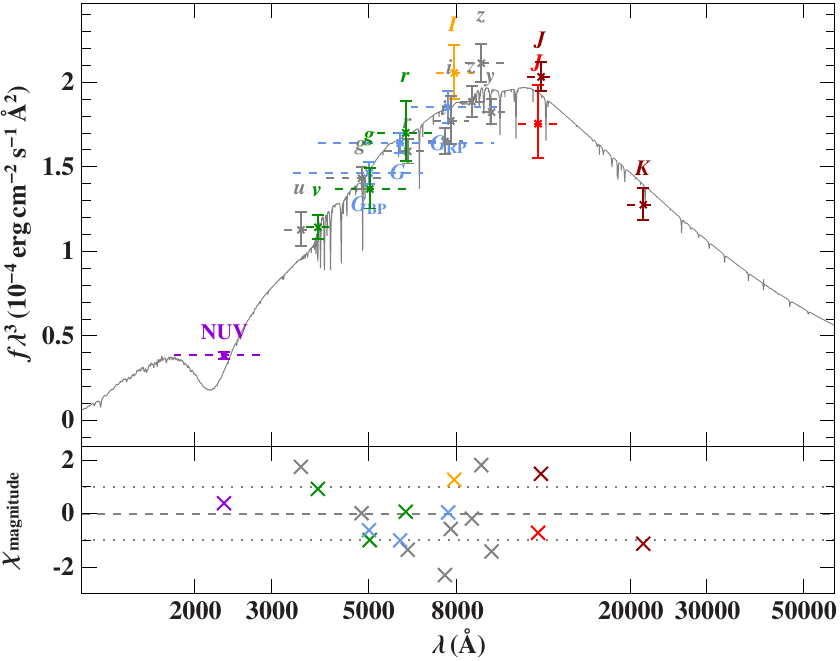}
    \caption{\small Best-fitting single-component model to the archival spectral energy distribution for \blg. Greyed data points were excluded from the fitting algorithm.}
    \label{fig:sed}
\end{figure}
We modeled the spectral energy distribution (SED) of \blg\ using model grids calculated with the ATLAS12 code \citep{kurucz1996} following the method described in \citet{heber2018}. We fit for the surface gravity, effective temperature, Helium abundance, angular diameter, and interstellar reddening. We included Gaussian priors on the surface gravity and effective temperature based on our spectroscopic modeling described above. Our best-fitting SED model is displayed in Figure \ref{fig:sed} including photometry from GALEX GR6$+$7 $NUV$ \citep{bianchi2017}, SkyMapper DR4 $v,g,r$ \citep{onken2024}, Gaia eDR3 $G_{\rm BP},G, G_{\rm RP}$ \citep{gaia2021}, DENIS $I$ \citep{denis2005}, 2MASS $J$ \citep{cutri2003}, and VISTA VHS DR6 $J,K$ \citep{mcmahon2013}.

We utilized the Gaia DR3 parallax ($\varpi=\parallax~\mas$), together with the fitted angular diameter, to obtain an estimate to the radius of the hot subdwarf donor in \blg. We find $R_{\rm sdOB}=\sedrvol~\rsol$, corresponding to mass $M_{\rm sdOB}=\sedmb~\msol$. This radius estimate is consistent with the spectroscopic radius obtained from the fitted rotational velocity with the assumption of spin-orbit synchronization, $R_{\rm sdOB,~v\sin{i}}=0.217^{+0.021}_{-0.009}~\rsol$, for orbital inclination ${\rm iangle}=79.7\pm0.4~{\rm deg}$, which was obtained from our eclipsing light curve fitting discussed in section \ref{sec:lcurve}.

\subsection{Light curve modeling}\label{sec:lcurve}
We performed simultaneous multi-band light curve modeling to our Bessel $B$, Bessel $R$, SDSS $g'$, and SDSS $i'$ light curves of \blg\ with a Markov Chain Monte Carlo approach using the \texttt{lcurve} \citep{copperwheat2010} software package, which models each flux-contributing component of the system as a dense grid of points emitting as blackbodies at a given temperature and central wavelength. Our light curve model included parameters for the hot subdwarf donor, the white dwarf accretor, and an irradiated accretion disc. 

We allowed the following model parameters to vary in our fit: mass ratio ($q=\frac{M_{\rm sdOB}}{M_{\rm WD}}$), orbital inclination (iangle), effective temperatures of the donor ($T_{\rm eff,~sdOB}$) and accretor ($T_{\rm eff,~WD}$), scaled radius of the accretor ($r_{\rm WD}=\frac{R_{\rm WD}}{a}$), velocity scale ($\frac{K_{\rm WD}+K_{\rm sdOB}}{\sin{i}}$), time of primary conjunction ($T_0$), scaled outer radius of the accretion disc ($r_{\rm 2,~disc}$), scaled height of the accretion disc ($h_{\rm disc}$), and the temperature of the accretion disc ($T_{\rm disc}$; a flux normalization parameter for the accretion disc). We additionally enabled the dataset-dependent flux trend parameter ``slope'' to remain free for each set of data. This slope parameter introduced a linear flux trend to the model and was used to account for color-dependent air mass effects.

We ran two sets of models to convergence: For solution A, we fixed the inner radius of the accretion disc to match the radius of the accretor (a common technique used for geometric simplification when modeling accretion disc in similar type binaries), applied filter-dependent gravity darkening and quadratic limb darkening values from \citet{claret2020}, and placed Gaussian priors on the temperature, surface gravity, velocity semi-amplitude, and the gravity darkening and quadratic limb darkening values based on our spectroscopic analysis. This allowed each of these quantities to vary, but penalized the model for straying far from the expected values estimated from our spectroscopic analysis. For solution B, we used an identical setup to solution A, but additionally included a Gaussian prior on the rotational velocity of the donor, which was constrained simultaneously with our spectroscopic surface gravity to model the observed broadened absorption lines. We present these two solutions separately to quantify the effects of including a prior on a rotational velocity which may be inaccurate due to unmodeled accretion disc flux distorting spectroscopic line profiles. The most-probable values for relevant fitted parameters are presented in Table \ref{tab:lcurve_table}. The most-probable values for all fitted parameters are presented in Appendix Table \ref{tab:lcurve_table_full}.

\begin{figure}
    \centering
    \includegraphics[width=1.0\linewidth]{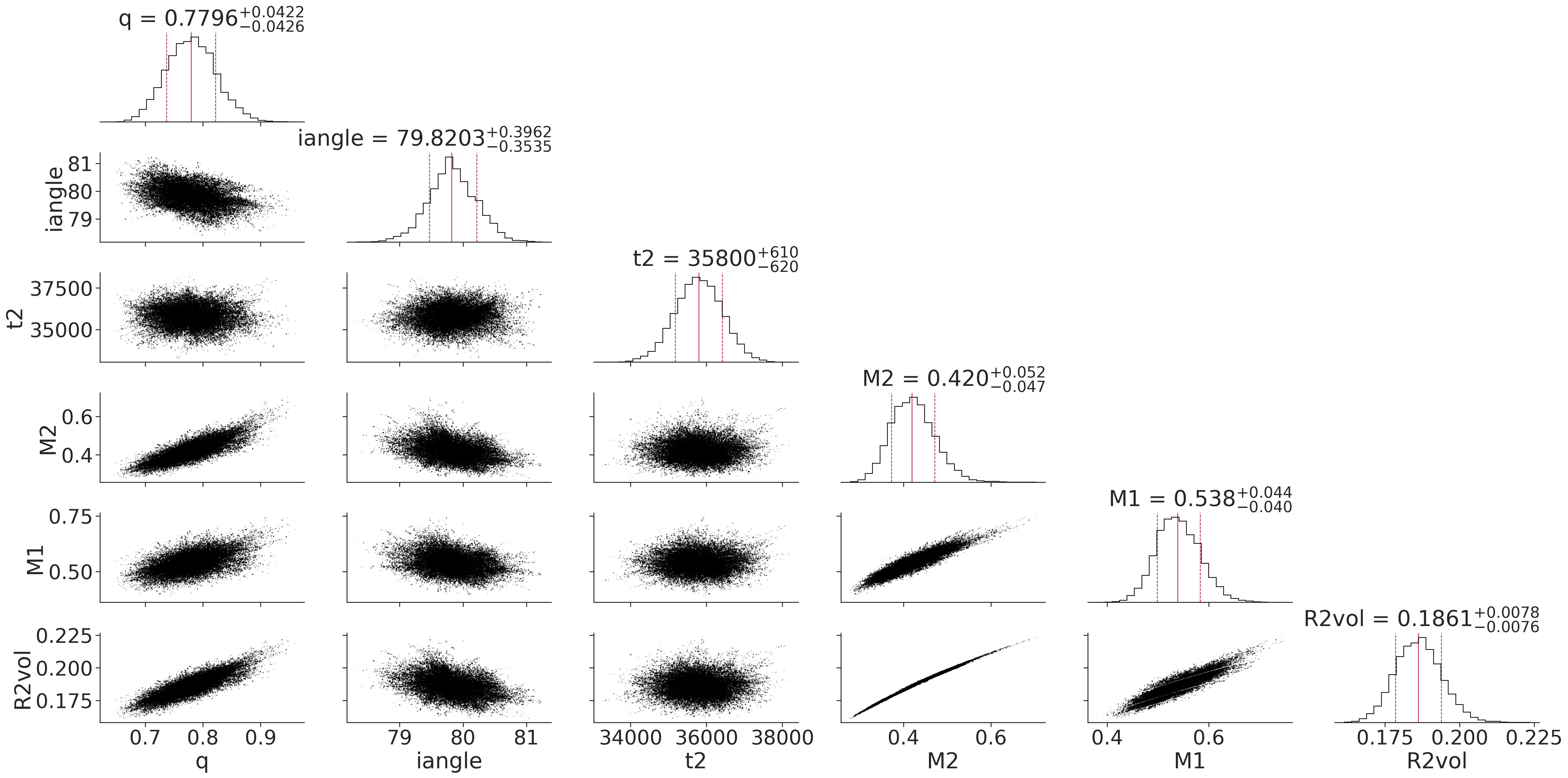}
    \caption{\small Posterior distributions for relevant system parameters of \blg\ derived from our MCMC chain output to our light curve modeling for Solution A as described in the text.}
    \label{fig:corner_plot}
\end{figure}

\begin{figure}
    \centering
    \includegraphics[width=0.95\linewidth]{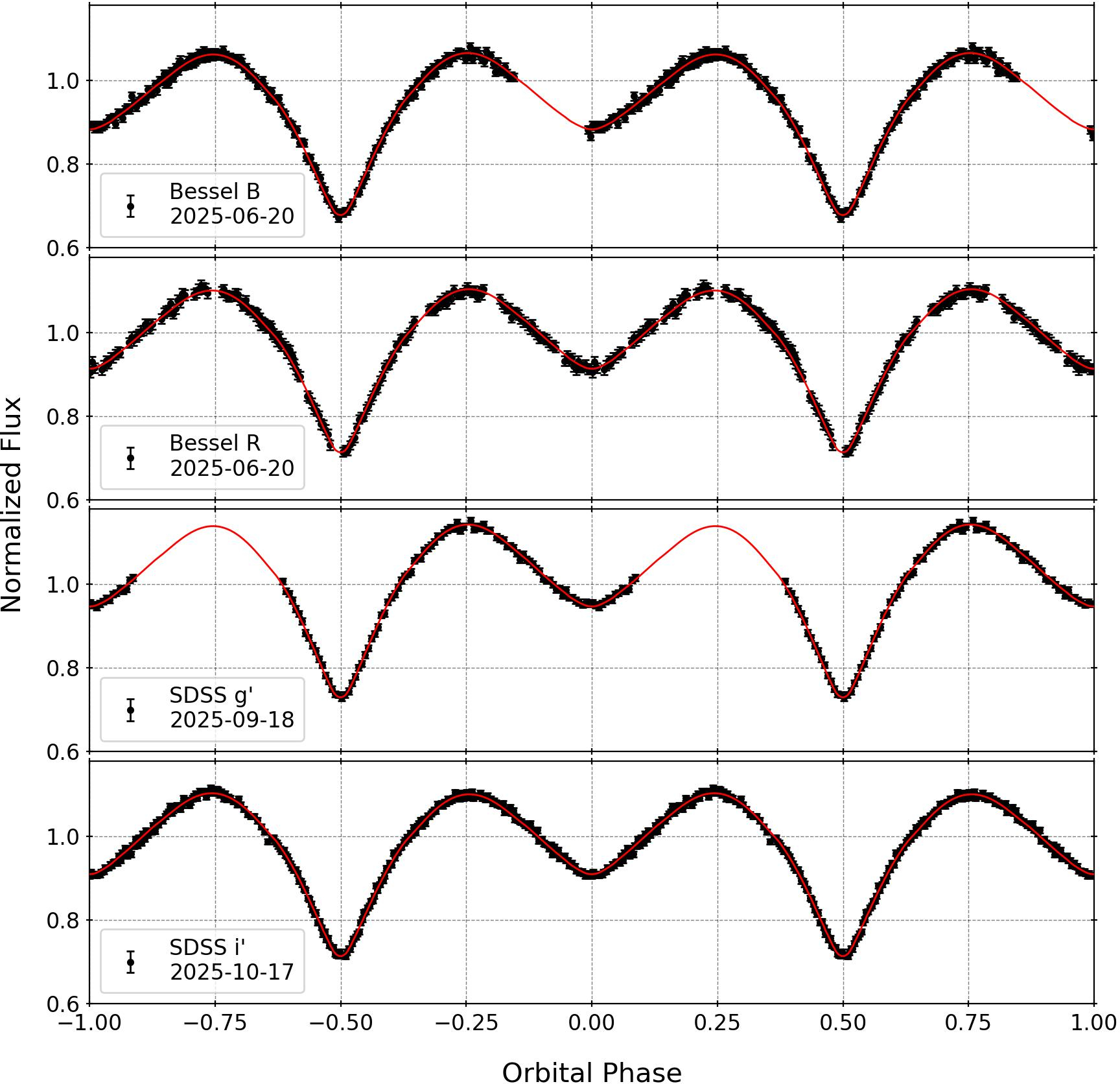}
    \caption{\small Best-fitting light curve model for Solution A (red line) over plotted onto our observed light curve data of \blg\ (black points) phase folded at the orbital period of the binary. Data has been repeated over one orbit for visual clarity.}
    \label{fig:lcurve_models}
\end{figure}

\begin{figure}
    \centering
    {\includegraphics[width=0.95\linewidth]{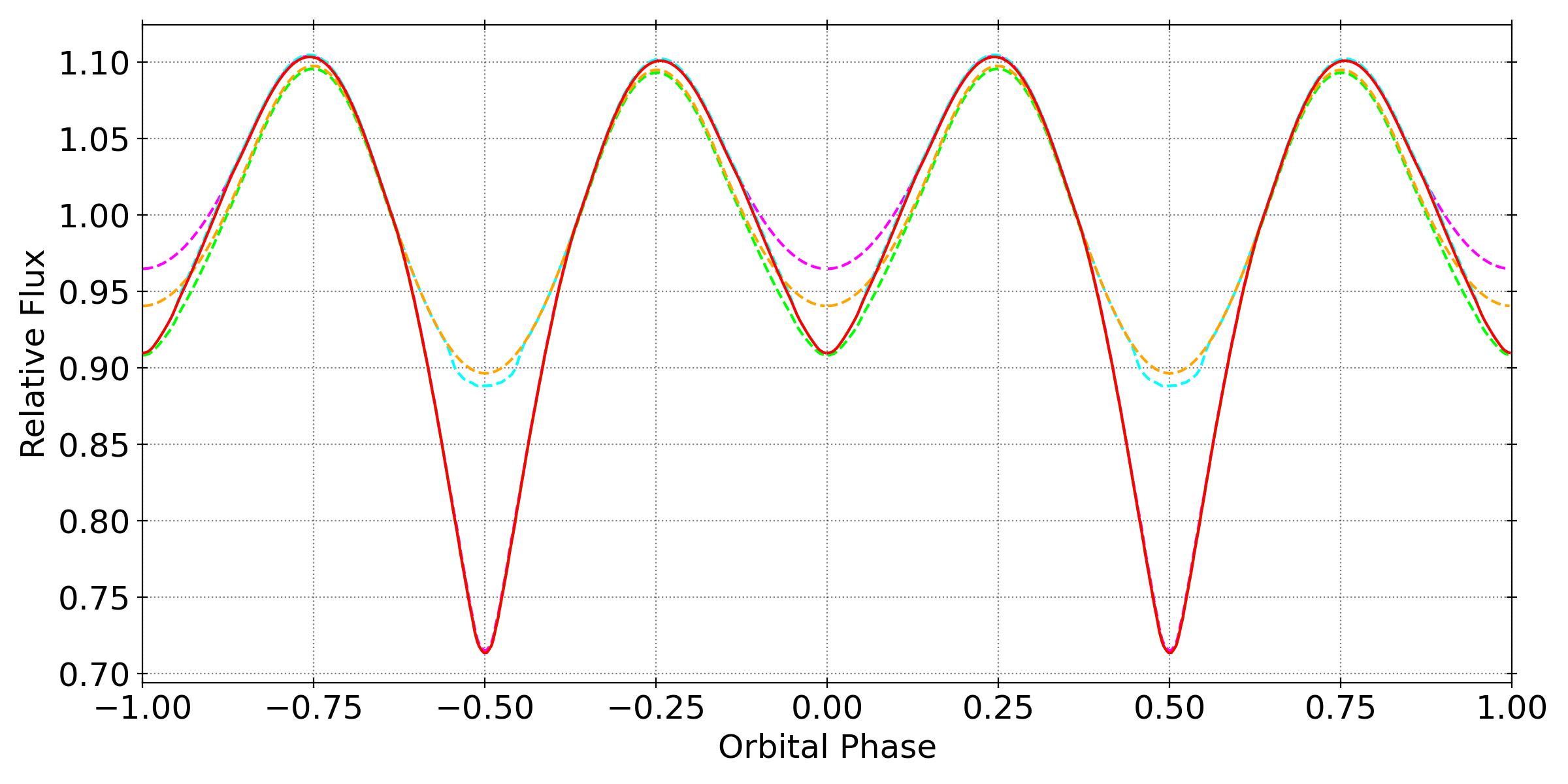}}
    \caption{\small Best-fitting light curve model (red) for \blg. Additional lines are included to demonstrate the effects of individually removing the irradiation of the outer accretion disc (green; no reprocessed irradiation from the accretion disc), enabling a transparent accretion disc (cyan; no eclipse of the sdOB donor by the accretion disc), and enabling a transparent subdwarf (magenta; no eclipse of the accretion disc by the sdOB donor). The orange line shows the model light curve without eclipses or disc irradiation effects and demonstrates the amplitude of tidal deformation of the donor without an accretion disc.}
    \label{fig:decomp_model}
\end{figure}

In both solutions, our models suggest that our data are not able to constrain the temperature or radius of the accreting white dwarf and that the shallow primary eclipse is an eclipse of the accretion disc, rather than an eclipse of the white dwarf. We present the posterior distributions for the relevant system parameters for \blg\ from solution A in Figure \ref{fig:corner_plot}. The most-probable model derived from solution A is over plotted onto the light curves of \blg\ in Figure \ref{fig:lcurve_models}. Solution B is visually identical to solution A when plotted together. We demonstrate the effects of removing flux-contributing components from the model in Figure \ref{fig:decomp_model}, which highlights the necessity of including a model component for an irradiated opaque accretion disc to obtain a reasonable fit to the observed light curves of \blg\ and similar systems.


\section{Discussion}
The complexity of binary light curve models with irradiated accretion discs highlights the difficulty in estimating accurate system parameters through light curve modeling alone, given the degeneracies between inclination, mass ratio, and reprocessed irradiation from the outer accretion disc.

Our light curve modeling solutions suggest that the accretion disc's flux contribution is about 4\% in the SDSS $i'$-band and about 2\% flux in the SDSS $g'$ band, which would not be explicitly visible in our optical spectroscopy given its noise level. While this contribution is small, it is still much larger than the expected accretion disc flux contribution in the similar Roche lobe-filling hot subdwarfs investigated by \citet{deshmukh2023}. Obtaining higher signal to noise spectroscopy of \blg\ may enable direct detection of its accretion disc to better characterize the binary.

\subsection{Roche-geometry constraints on donor mass}
We estimated the mean density of the hot subdwarf donor star in \blg\ using the relationship between the binary orbital period and binary mass ratio for Roche lobe-filling binaries \citep{eggleton1983}. This density estimate is weakly dependent on mass ratio and provides a separate verification to the mass of the donor when combined with an independent and robust radius estimate, such as from an eclipsing light curve model or an SED fit with a precise distance estimate. Because our eclipsing light curve model radius is degenerate with the disc geometry, we make use of the radius from our SED fit for this calculation, but note that our SED fit included only one stellar component without a disc.
 \begin{table}
    \renewcommand{\arraystretch}{1.5}
    \small
    \centering
    \caption{Most-probable binary parameters for our two presented solutions based on our light curve modeling of \blg.}
    \label{tab:lcurve_table}
    
    \begin{tabular*}{0.99\columnwidth}{@{\extracolsep{\fill}}l r r}
        \hline\hline
        {} & {Solution A} & {Solution B} \\
        \hline
            $q=\frac{M_{\rm sdOB}}{M_{\rm WD}}$   & $\lcurveq$      & $\lcurveqB$ \\
            ${\rm iangle}$ ($\deg$)              & $\lcurveiangle$ & $\lcurveiangleB$ \\
            $T_{\rm eff,~sdOB}$ ($\kelvin$)       & $\lcurvetsdb$   & $\lcurvetsdbB$ \\
            $R_{\rm sdOB,~volumetric}$ ($\rsol$)  & $\lcurvervol$   & $\lcurvervolB$ \\
            $M_{\rm sdOB}$ ($\msol$)              & $\lcurvemb$     & $\lcurvembB$ \\
            $M_{\rm WD}$ ($\msol$)               & $\lcurvema$     & $\lcurvemaB$ \\[0.1cm]
            $a$ ($\rsol$) & $\lcurvea$   & $\lcurveaB$ \\[0.1cm]
            $\left<\rho_{\rm sdOB}\right>_{\rm Eggleton1983}$ ($\rhocgs$) & $\eggletondensity$ & $\eggletondensityB$ \\[0.1cm]
        \hline
        \hline
    \end{tabular*}
    \raggedright
    \footnotesize
    Notes: We include the calculated binary separation and an estimate of the mean density of the donor using the period- density relation from \citet{eggleton1983} for each solution.
\end{table}

We find expected mean donor densities $\left<\rho_{\rm sdOB}\right>_A=\eggletondensity~\rhocgs$ and $\left<\rho_{\rm sdOB}\right>_B=\eggletondensityB~\rhocgs$ using the mass ratios from our two light curve model solutions. These both correspond to donor mass $M_{\rm sdOB,~Eggleton}=\eggletonmass~\msol$ when combined with our best fitting volumetric radius from our SED modeling. This mass estimate does not provide strong constraints due to uncertainties in the radius estimate from our SED fit.

\subsection{Orbital period change estimates}
Long term photometric monitoring for changes in the orbital period of \blg\ may enable constraints on the chirp mass of the binary. An orbital decay analysis of the Roche lobe-filling sdOB+WD binary ZTF J2130+4420 has been completed by \citet{teckenburg2025}, who find a chirp mass estimate consistent with the estimate from the photometric masses presented in \citet{kupfer2020a} based on light curve modeling.

To measure the observed period change in \blg, we made use of the ATLAS forced photometry service \citep{shingles2021}. We used the reduced image forced photometry data of \blg\ spanning 2015 June to 2025 August, grouping the data into bins with width two years. This provided 5 bins with [453, 592, 950, 1057, 1065] data points. We converted the ATLAS timings from start of exposure MJD\_UTC to mid-exposure BJD\_TDB based on the ATLAS observatory's site ID found in the archival light curve data file.

We estimated the time of primary conjunction, $T_0$, for each data bin by modeling each subset of data using an MCMC implementation of \texttt{lcurve}, fitting only for the time of primary conjunction, $T_0$, while keeping all other parameters fixed to our best-fitting parameters for Solution A, described in Section \ref{sec:photo_analysis}. We initialized the walkers around the expected ephemeris nearest the mid-point date of each data bin, assuming constant period. The resulting ephemeris values from our 1-dimensional MCMC chain for each data bin are presented in Table \ref{tab:pdot}. We used these values to attempt an (O-C) analysis to infer the change in orbital period of \blg, but find $\dot{P}=(\pdotoc)\times10^{-12}~{\rm s~s^{-1}}$, consistent with no measurable period change within the 10-year baseline provided by ATLAS.
 \begin{table}
    \renewcommand{\arraystretch}{1.5}
    \small
    \centering
    \caption{\small Best-fitting mid-eclipse timings and their associated $1\sigma$ confidence ranges for \blg.}
    \label{tab:pdot}
    \begin{tabular*}{0.99\columnwidth}{@{\extracolsep{\fill}}c c c}
        \hline\hline
        {$T_0$} & {$\sigma_{T_0}$} & {Facility} \\
        {${\rm BJD_{TDB}}$} & {${\rm BJD_{TDB}}$} & {} \\
        {(${\rm d}$)} & {(${\rm s}$)} & {} \\
        \hline
        2457720.9673878 & $^{+5.9}_{-5.8}$ & ATLAS \\[0.2cm]
        2458578.1057461 & $^{+4.6}_{-4.5}$ & ATLAS \\[0.2cm]
        2459308.0730657 & $^{+2.7}_{-3.3}$ & ATLAS \\[0.2cm]
        2460038.0850321 & $^{+3.6}_{-3.6}$ & ATLAS \\[0.2cm]
        2460768.0966661 & $^{+3.8}_{-4.0}$ & ATLAS \\[0.2cm]
        2460936.6224013 & $^{+0.8}_{-0.8}$ & SOAR \\[0.2cm]
        \hline
        \hline
    \end{tabular*}
\end{table}

Based on the system parameters derived from our light curve modeling, we estimated the expected orbital period change due to gravitational wave radiation to be $(\pdot)\times10^{-13}~{\rm s~s^{-1}}$ and $(\pdotB)\times10^{-13}~{\rm s~s^{-1}}$ for solutions A and B, respectively. While neither of these solutions is favored over the other by our observational constraints, we note that an orbital period change on the order of solution B may become measurable with continued monitoring of \blg\ over the next few years. However, the active mass transfer in \blg\ is expected to widen its orbit at a rate determined by the donor's response to mass loss and the mass transfer rate. We find that an instantaneous mass transfer rate of $\dot{M}_{\rm ~critical}\approx(5-10)\times10^{-9}~{\rm M_\odot~yr^{-1}}$ would be necessary to widen the orbit in balance with the instantaneous orbital decay from gravitational wave emission.

\subsection{Future evolution}

Our two solutions for \blg\ are consistent with a broad range of possible masses for the hot subdwarf donor, with both solutions consistent with \blg\ containing a canonical mass hot subdwarf. Because of this, it is difficult to precisely determine the evolutionary history of \blg\ based on our modeling solutions.
Obtaining higher quality spectroscopic data of \blg\ and revisiting its modeling with a two-component fit, including flux from an accretion disc, would improve the spectroscopic parameter estimates and potentially enable a unique light curve solution.

Comparing our light curve solutions to the literature sample of four previously studied Roche lobe-filling hot subdwarfs, we note that the mass of \blg\ derived from solution A ($M_{\rm A}=\lcurvemb~\msol$) more closely matches the observed masses of ZTF J2130$+$4420 ($M_{\rm sdOB}=0.337\pm0.015~\msol$), ZTF J2055$+$4651 ($M_{\rm sdOB}=0.041\pm0.04~\msol$), and ZTF J0007$+$4804 ($M_{\rm sdB}=0.42\pm0.01~\msol$), while the mass derived from solution B ($M_{\rm B}=\lcurvembB~\msol$) more closely matches SMSS J1920-2001 ($M_{\rm sdO}=0.55~\msol$). 

With the exception of ZTF J0007$+$4804, the other known Roche-lobe filling hot subdwarfs are expected to be undergoing shell-burning, with remaining shell-burning lifetimes of $\mathcal{O}(1~\myr)$. This is much shorter than the expected orbital evolutionary timescale of \blg\ from the effects of gravitational wave radiation and mass transfer. In this shell-burning scenario, both of our light curve solutions suggest that the hot subdwarf donor in \blg\ will evolve into a low mass CO-core white dwarf, forming a detached double white dwarf binary that will once again be driven to interaction by gravitational wave emission, with evolutionary timescale on the order of $\tau_{\rm GW}\approx\mathcal{O}(10~{\rm Myr})$. The eventual fate of \blg\ then depends on its mass ratio and accretor core composition. Mass transfer as a double white dwarf binary may result in a double-detonation supernova explosion \citep{shen2024}, a direct merger into a single massive white dwarf, or form a stable AM CVn type accreting binary.

\vspace{0.1cm}
\section*{Acknowledgments}
Based on observations obtained at the Southern Astrophysical Research (SOAR) telescope, which is a joint project of the Ministério da Ciência, Tecnologia e Inovações do Brasil (MCTI/LNA), the US National Science Foundation’s NOIRLab, the University of North Carolina at Chapel Hill (UNC), and Michigan State University (MSU).

We are grateful to the SOAR support staff for their invaluable help during our observations, in particular Sergio Pizarro, Juan Espinoza, Carlos Corco, and Fernando Cort\'{e}s.

We thank the anonymous referee for the comments and suggestions that improved the quality of this work.

This publication makes use of data products from the Two Micron All Sky Survey, which is a joint project of the University of Massachusetts and the Infrared Processing and Analysis Center/California Institute of Technology, funded by the National Aeronautics and Space Administration and the National Science Foundation.

The national facility capability for SkyMapper has been funded through ARC LIEF grant LE130100104 from the Australian Research Council, awarded to the University of Sydney, the Australian National University, Swinburne University of Technology, the University of Queensland, the University of Western Australia, the University of Melbourne, Curtin University of Technology, Monash University and the Australian Astronomical Observatory. SkyMapper is owned and operated by The Australian National University's Research School of Astronomy and Astrophysics. The survey data were processed and provided by the SkyMapper Team at ANU. The SkyMapper node of the All-Sky Virtual Observatory (ASVO) is hosted at the National Computational Infrastructure (NCI). Development and support of the SkyMapper node of the ASVO has been funded in part by Astronomy Australia Limited (AAL) and the Australian Government through the Commonwealth's Education Investment Fund (EIF) and National Collaborative Research Infrastructure Strategy (NCRIS), particularly the National eResearch Collaboration Tools and Resources (NeCTAR) and the Australian National Data Service Projects (ANDS).

The VISTA Hemisphere Survey data products served at Astro Data Lab are based on observations collected at the European Organisation for Astronomical Research in the Southern Hemisphere under ESO programme 179.A-2010, and/or data products created thereof.

This work has made use of data from the Asteroid Terrestrial-impact Last Alert System (ATLAS) project. The Asteroid Terrestrial-impact Last Alert System (ATLAS) project is primarily funded to search for near earth asteroids through NASA grants NN12AR55G, 80NSSC18K0284, and 80NSSC18K1575; byproducts of the NEO search include images and catalogs from the survey area. This work was partially funded by Kepler/K2 grant J1944/80NSSC19K0112 and HST GO-15889, and STFC grants ST/T000198/1 and ST/S006109/1. The ATLAS science products have been made possible through the contributions of the University of Hawaii Institute for Astronomy, the Queen’s University Belfast, the Space Telescope Science Institute, the South African Astronomical Observatory, and The Millennium Institute of Astrophysics (MAS), Chile.

This work made use of Astropy:\footnote{http://www.astropy.org} a community-developed core Python package and an ecosystem of tools and resources for astronomy \citep{astropy2013, astropy2018, astropy2022}. 

This research made use of the High Performance Computing Resources at the University of North Carolina at Chapel Hill.

\bibliographystyle{aasjournal}
\bibliography{00_Main.bib}

\begin{appendix}
\renewcommand{\thefigure}{A\arabic{figure}}
\renewcommand{\theHfigure}{A\arabic{figure}}
\setcounter{figure}{0}

\renewcommand{\thetable}{A\arabic{table}}
\renewcommand{\theHtable}{A\arabic{table}}
\setcounter{table}{0}

\section{Supplemental Figures and Tables}\label{sec:appendixA}
 \begin{table}[H]
    \renewcommand{\arraystretch}{1.5}
    \small
    \centering
    \caption{\small Most-probable fitted parameters for our two solutions based on our light curve modeling with \texttt{lcurve}.}
    \label{tab:lcurve_table_full}
    \begin{tabular*}{0.99\columnwidth}{@{\extracolsep{\fill}}l l r r}
        \hline\hline
        {Parameter Name} & {Description} & {Solution A} & {Solution B} \\
        \hline
            \texttt{q} & Mass ratio $\frac{M_{\rm sdOB}}{M_{\rm WD}}$ & $\lcurveq$      & $\lcurveqB$ \\
            \texttt{iangle} & Orbital inclination ($\deg$) & $\lcurveiangle$ & $\lcurveiangleB$ \\[0.3em]
            \texttt{r1} & Scaled primary radius $\frac{R_{WD}}{a}$ & $\lcurverwd$ & $\lcurverwdB$ \\[0.3em]
            \texttt{t1} & $T_{\rm eff,~WD}$ ($\kelvin$) & $\lcurvetwd$   & $\lcurvetwdB$ \\[0.3em]
            \texttt{t2} & $T_{\rm eff,~sdOB}$ ($\kelvin$) & $\lcurvetsdb$ & $\lcurvetsdbB$ \\[0.3em]
            \texttt{velocity\_scale} & $\frac{K_{\rm sdOB}+K_{\rm WD}}{\sin{\left(\rm iangle\right)}}$ (${\rm km~s^{-1}}$) & $\lcurvevscale$   & $\lcurvevscaleB$ \\[0.3em]
            \texttt{t0} & $T_0$ (${\rm BJD\_TDB~d\pm s}$) & $\lcurvemidtime$ & $\lcurvemidtimeB$\\[0.5em]
            \texttt{rdisc2} & $\frac{R_{\rm disc,~outer}}{a}$ & $\lcurverdisc$ & $\lcurverdiscB$\\[0.3em]
            \texttt{height\_disc} & $\frac{h_{1/2,~disc}}{a}$ & $\lcurvehdisc$ & $\lcurvehdiscB$\\[0.3em]
            \texttt{temp\_disc} & $T_{\rm disc}$ ($\kelvin$) & $\lcurvetdisc$ & $\lcurvetdiscB$\\[0.3em]
            \texttt{temp\_edge} & $T_{\rm disc,~outer~edge}$ ($\kelvin$) & $\lcurvetempedge$ & $\lcurvetempedgeB$\\[0.5em]
            \hline
            \texttt{slope\_B} & Linear light curve flux trend (B-band) & $\lcurveslopeB$ & $\lcurveslopeBB$\\
            \texttt{ldc2\_1\_B} & \citet{claret2020} coefficient a & $\lcurveldcAB$ & $\lcurveldcABB$\\
            \texttt{ldc2\_2\_B} & \citet{claret2020} coefficient b & $\lcurveldcBB$ & $\lcurveldcBBB$\\
            \texttt{gravity\_dark2\_B} & \citet{claret2020} ($\beta_1=1.0$) & $\lcurvegdcB$ & $\lcurvegdcBB$\\[0.5em]
            \hline
            \texttt{slope\_R} & Linear light curve flux trend (R-band) & $\lcurveslopeR$ & $\lcurveslopeRB$\\
            \texttt{ldc2\_1\_R} & \citet{claret2020} coefficient a & $\lcurveldcAR$ & $\lcurveldcARB$\\
            \texttt{ldc2\_2\_R} & \citet{claret2020} coefficient b & $\lcurveldcBR$ & $\lcurveldcBRB$\\
            \texttt{gravity\_dark2\_R} & \citet{claret2020} ($\beta_1=1.0$) & $\lcurvegdcR$ & $\lcurvegdcRB$\\[0.5em]
            \hline
            \texttt{slope\_g} & Linear light curve flux trend (g-band) & $\lcurveslopeg$ & $\lcurveslopegB$\\
            \texttt{ldc2\_1\_g} & \citet{claret2020} coefficient a & $\lcurveldcAg$ & $\lcurveldcAgB$\\
            \texttt{ldc2\_2\_g} & \citet{claret2020} coefficient b & $\lcurveldcBg$ & $\lcurveldcBgB$\\
            \texttt{gravity\_dark2\_g} & \citet{claret2020} ($\beta_1=1.0$) & $\lcurvegdcg$ & $\lcurvegdcgB$\\[0.5em]
            \hline
            \texttt{slope\_i} & Linear light curve flux trend (i-band) & $\lcurveslopei$ & $\lcurveslopeiB$\\
            \texttt{ldc2\_1\_i} & \citet{claret2020} coefficient a & $\lcurveldcAi$ & $\lcurveldcAiB$\\
            \texttt{ldc2\_2\_i} & \citet{claret2020} coefficient b & $\lcurveldcBi$ & $\lcurveldcBiB$\\
            \texttt{gravity\_dark2\_i} & \citet{claret2020} ($\beta_1=1.0$) & $\lcurvegdci$ & $\lcurvegdciB$\\[0.5em]
        \hline
        \hline
    \end{tabular*}
\end{table}
\begin{figure*}
    \centering
    \includegraphics[width=1.0\linewidth]{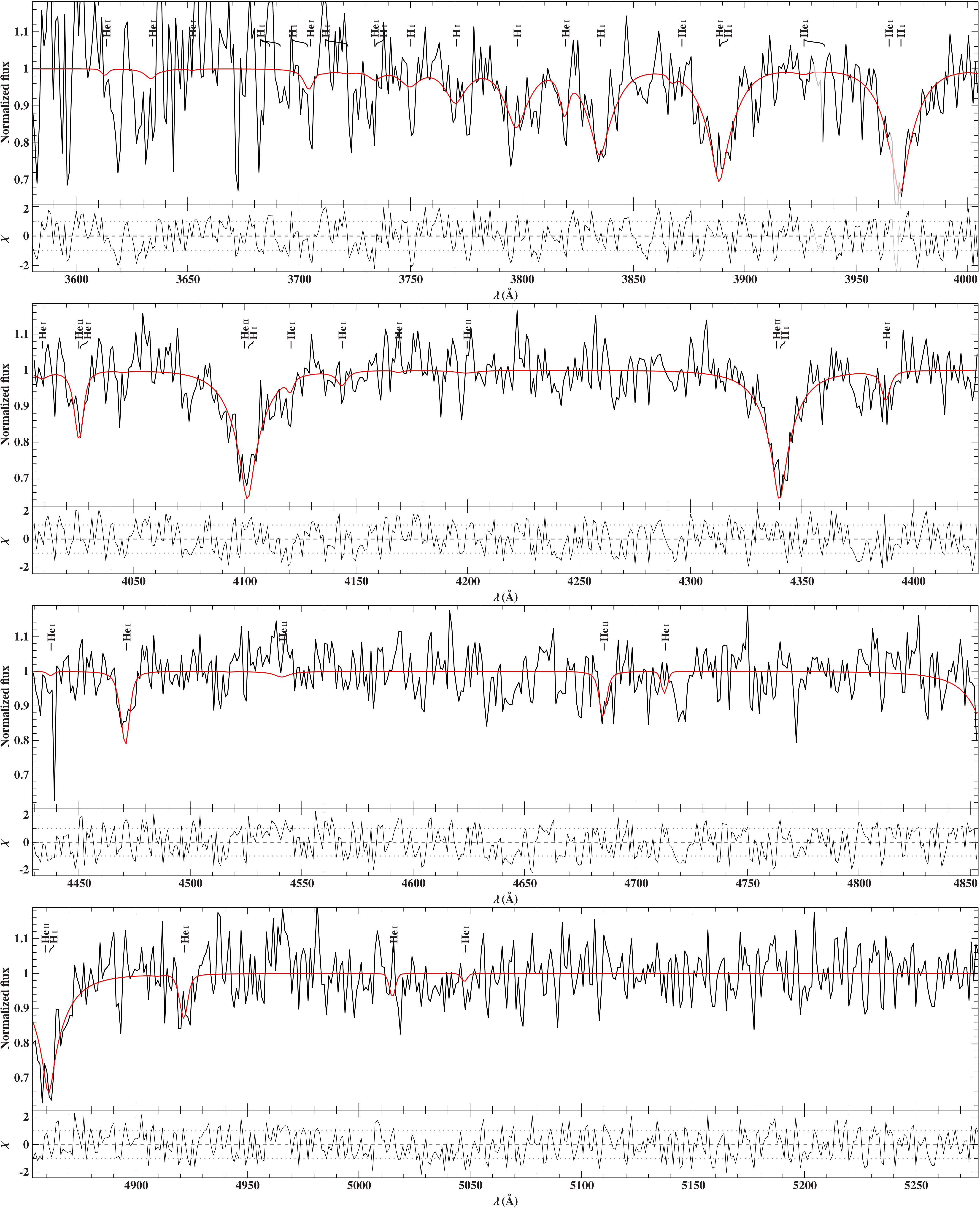}
    \caption{\small Best fitting model (red) overplotted onto our observed SOAR optical spectrum of \blg\ (black) for orbital phase $\phi=0.0$. Common features for Hydrogen and Helium are marked. Greyed regions represent data excluded from the fit, including absorption from the Ca II H \& K doublet.}
    \label{fig:spectrum_fit00}
\end{figure*}

\begin{figure*}
    \centering
    \includegraphics[width=1.0\linewidth]{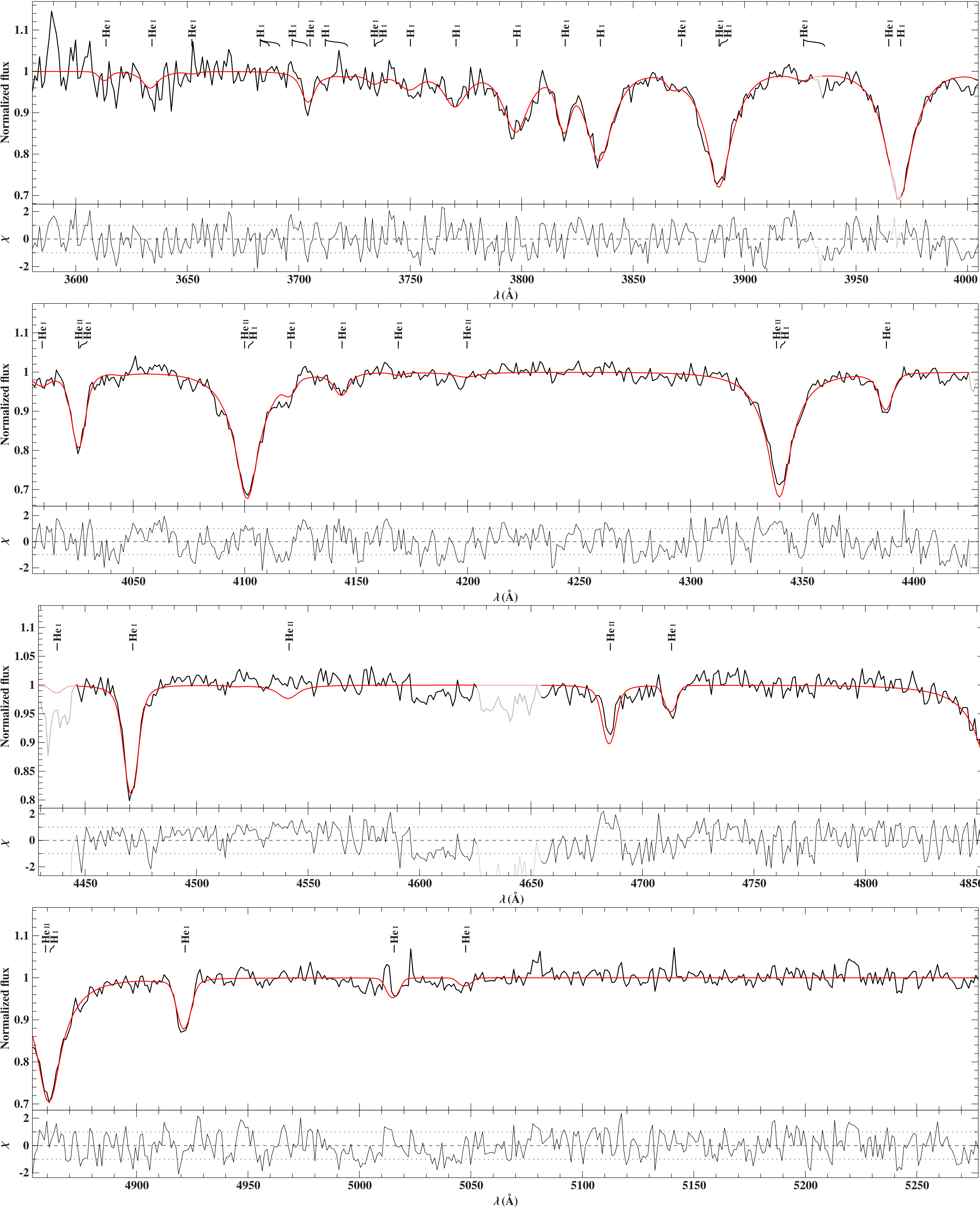}
    \caption{\small Best fitting model (red) over plotted on to our zero-velocity, combined, optical spectrum of \blg\ (black). Common features for Hydrogen and Helium are marked. Greyed regions represent data excluded from the fit, including absorption from the Ca II H \& K doublet.}
    \label{fig:spectrum_fit}
\end{figure*}
\end{appendix}

\end{document}